\documentclass[aps,twocolumn]{revtex4-1}
\usepackage{amsmath,amssymb}
\usepackage{slashed}
\usepackage{graphicx}
\usepackage{bm}
\usepackage[top=1.0in,bottom=1.0in,left=1.0in,right=1.0in]{geometry}
\usepackage[colorlinks,linkcolor=blue,citecolor=blue]{hyperref}

\newcommand{\be}{\begin{equation}}
\newcommand{\ee}{\end{equation}}
\newcommand{\bear}{\begin{eqnarray}}
\newcommand{\eear}{\end{eqnarray}}
\newcommand{\ba}{\begin{array}}
\newcommand{\ea}{\end{array}}

\def\be{\begin{eqnarray}}
\def\ee{\end{eqnarray}}
\def\bea{\be}
\def\eea{\ee}

\def\roughly#1{\mathrel{\raise.3ex\hbox{$#1$\kern-.75em%
\lower1ex\hbox{$\sim$}}}}

\begin{document}

\title{Hyperons and $\Theta_s^+$ in Holographic QCD}
%\title{Hunting pentaquarks in photo-production of heavy mesons}

%\author{Arthur Kock}
%\email{arthur.kock@stonybrook.edu}
%\affiliation{Center for Nuclear Theory, Department of Physics and Astronomy, Stony Brook University, Stony Brook, New York 11794--3800, USA}

\author{Yizhuang Liu}
\email{yizhuang.liu@uj.edu.pl}
\affiliation{Institute of Theoretical Physics,
Jagiellonian University, 30-348 Kraków, Poland}

\author{Maciej A. Nowak}
\email{maciej.a.nowak@uj.edu.pl}
\affiliation{Institute of Theoretical Physics and Mark Kac Center for Complex Systems Research,
Jagiellonian University, 30-348 Kraków, Poland}

\author{Ismail Zahed}
\email{ismail.zahed@stonybrook.edu}
\affiliation{Center for Nuclear Theory, Department of Physics and Astronomy, Stony Brook University, Stony Brook, New York 11794--3800, USA}

%\author{
%Yizhuang Liu and Ismail Zahed }

%\affiliation{ Department of Physics and Astronomy, \\ Stony Brook University,\\
%Stony Brook, NY 11794, USA}

\begin{abstract}
We  revisit the holographic description of strange baryons in the context of the Sakai-Sugimoto construction,
by considering  the strange quark mass as heavy. Hyperons are described by a  massive $(K,K^*)$ multiplet,
bound to a light-flavor instanton in bulk, much in the spirit of the  Callan-Klebanov construction.
 The modular Hamiltonian maps onto the Landau problem,
a charged particle in a 2-dimensional external magnetic
field, induced by the bulk Chern-Simons interaction, plus spin-orbit coupling. The ensuing holographic hyperon spectrum compares fairly with  the  empirical one.
The  holographic strange pentaquark baryon $\Theta^+_s$ is shown to be unbound.
%This approach is tthe holographic analogue of the Callan-Klebanov aconstruction for hyperons.
\end{abstract}

\maketitle

%%%%%%%%%%%%%%%%%%%%%%%%%%%%%%%%%%%%%%%%

\section{Introduction}

The holographic principle in general~\cite{Maldacena:1997re,Erlich:2005qh},
and the D4-D8-D$\bar 8$ holographic set-up in particular~\cite{Sakai:2004cn} provide a framework for addressing QCD in the
infrared in the double limit of a large number of colors and strong $^\prime$t Hooft gauge coupling $\lambda=g_{YM}^2N_c$. It is
confining and exhibits spontaneous  chiral symmetry breaking geometrically. The light meson sector is well described by an effective action
with manifest chiral symmetry and very few parameters, yet totally
in line with  more elaborate  effective theories of QCD~\cite{Fujiwara:1984mp}. The same set-up can be minimally
modified to account for the description of heavy-light mesons, with manifest  heavy
quark symmetry~\cite{Liu:2016iqo,Liu:2017xzo,Liu:2017frj,Li:2017dml,Fujii:2020jre}.

Light and heavy-light baryons are dual to instantons and instanton-heavy meson bound states in
bulk~\cite{Hata:2007mb,Hashimoto:2008zw,Kim:2008pw,Hata:2007tn,Hashimoto:2009st,Lau:2016dxk}, providing a
robust geometrical  approach to the multi-body bound state problem.
The holographic construction provides a  dual realization of  the chiral soliton approach and its bound states variants~\cite{Zahed:1986qz,Rho:1992yy},
without the shortcomings of the derivative expansion. It is a geometrical  realization of the molecular  approach
~\cite{Wu:2010jy,Karliner:2015ina}, without the ambiguities of the nature of the meson exchanges, and the arbitrariness in the choice of the
many couplings and form factors~\cite{Lin:2019qiv}. Alternative holographic models for the description of heavy hadrons, have been
developed  in~\cite{Dosch:2015bca,Sonnenschein:2018fph}.

Chiral symmetry constrains the light quark interactions, while heavy quark symmetry
restricts the spin interactions between heavy quarks~\cite{Shuryak:1981fza,Isgur:1991wq}.
Both symmetries are inter-twined by the phenomenon  of chiral
doubling~\cite{Nowak:1992um,Bardeen:1993ae,Nowak:2004jg} as shown experimentally in~\cite{Aubert:2003fg,Besson:2003cp}.
A theoretical approach to the multiquark states should have manifest chiral and heavy quark symmetry, a clear organizational
principle in the confining regime, and should address concisely the multi-body bound state problem.  The holographic construction
provides this framework.

In~\cite{Liu:2017frj} two of us have analyzed the holographic  baryon spectrum by considering three massless lavors $u,d,s$.
The  strange quark mass was introduced through a bulk instanton holonomy, assumed small and treated in perturbation theory.
However, the strange quark mass is intermediate between heavy and light, and may require a treatment beyond
perturbation theory. In this work, we will propose such a treatment.

We will consider  the kaon mass as large,  and identify the strangeness brane as heavy
in  the formulation outlined in~\cite{Liu:2017xzo,Liu:2017frj,Liu:2021tpq}. Hyperons (baryons with strangeness $-1$)
will be sought as bound states of massive kaons,  to a bulk flavor instanton made of only the light $u,d$ flavors.
The ensuing modular Hamiltonian for the hyperons,  will be diagonalized without recourse to
perturbation theory. This construction bears much in common with the revised bound state
approach for hyperons in the Skyrme model~\cite{Callan:1985hy,Callan:1987xt}. The result is a much  improved holographic hyperon spectrum,
in comparison to the one in~\cite{Liu:2017frj}.

The organization of the paper is as follows: In section~\ref{sec_modular}, we detail the modular  Lagrangian
for hyperons (baryons with strangeness $-1$)
in leading order in the heavy meson mass expansion. We retain  the exact kaon mass contribution, and the
non-local contributions stemming from the Coulomb back-reaction and the bulk form of Gauss law.
In section~\ref{sec_spectrum1} we show that the modular Hamiltonian maps onto the Landau
problem of a particle in a magnetic field in 2-dimensions, plus spin-orbit coupling.  In section~\ref{sec_spectrum2}
we detail the hyperon spectra, including  the strange  pentaquark exotic $\Theta_s^+$,
for two different approximations of the Gauss law contribution. Our conclusions are in~\ref{sec_conclusions}.
A number of Appendices are added to complement the derivations in the text.

%%%%%%%%%%%%%%%%%%%%%%%%%%%%%%%%%%%%%%%%
\section{The Modular Lagrangian}~\label{sec_modular}
% the total Lagrangian can now be calculated as by combing with other contributions in
%We first collect all the quadratic and quartic terms of the Lagrangian.
The modular Lagrangian for the holographic description of heavy-light mesons bound to a bulk flavor instanton, has
been discussed in~\cite{Liu:2017xzo,Liu:2017frj} for standard baryons, and for their exotics
in~\cite{Liu:2019mxw,Liu:2019yye,Liu:2021tpq,Liu:2021ixf}. Here we propose to use it for kaons,
assuming the strangeness to be a heavy flavor.

In brief, we identify the strange heavy-light flavor field $\Phi$ in bulk with  the $(0^-,1^-)$ kaon multiplet
 as $\Phi=(K,K^*)$, and proceed to bind it to a light flavor instanton as in~\cite{Liu:2017xzo,Liu:2021tpq}.
The ensuing modular Lagrangian is composed of the collective variables
$(X_i, a_I, \rho)$ for the instanton collective position, SU(2)  orientation and size.

In addition, when  binding to the core instanton in bulk, the kaon multiplet transmutes to
 a 2-component complex modular coordinate $\chi$, via a zero-mode. In the
 analysis to follow for the hyperon spectrum, this coordinate will be quantized as
 a boson, much in the spirit  of the bound state approach
 in the dual analysis in~\cite{Callan:1985hy,Callan:1987xt}.
 The fermionic statistics was considered in~\cite{Liu:2017xzo,Liu:2021tpq},
 in the analysis of the much heavier baryons and their exotics,  as it  captures the key
 features of the heavier quark, and heavy quark symmetry.

\subsection{General}

The full holographic modular Lagragian for heavy-light kaons bound to a bulk flavor instanton,
is given by (see Eq. 23 in~\cite{Liu:2021tpq})

\begin{align}
\label{FULL}
{\cal L}=&+\frac{1}{2}\dot \chi^{\dagger}\dot \chi+\frac{3i}{\tilde \rho^2}\chi^{\dagger}\dot \chi-\frac{37+12\frac{Z^2}{\rho^2}}{192}\chi^{\dagger}\chi\nonumber\\
&+\frac{78i}{5\tilde \rho^2}\chi^{\dagger}\tau^a\chi \chi^a -\frac{12}{5\tilde \rho^4} (\chi^\dagger \tau^a\chi)^2 \nonumber \\
&+\left(\frac{1}{4 }\frac{\dot\rho^2}{\rho^2}+\frac{ \dot a_I^2}{4}+\frac{\dot X^2}{4\rho^2}\right)\chi^{\dagger}\chi-\frac{1}{2}m_H^2\chi^{\dagger}\chi
\nonumber\\
&+{\cal L}_{\Phi_0} [m_H] +{\cal L}_{\rm Coulomb}\,,
\end{align}
where $\rho$ is the size of the instanton with $\tilde \rho=16\pi^2a N_c \rho$. The rescaling $\chi\rightarrow e^{im_Ht}\sqrt{m_H}\chi$
with the heavy and bare mass $m_H$ of the $(K,K^*)$ multiplet,   subsumed.
The moduli of $SU(2)$ rotation reads $$\chi^a=Tr(\tau^a{a}_I^{-1}\dot{a}_I)\ .$$

The first two lines are standard,
with the first term in the third line following from the coupling ${\rm tr } \Phi^2 \chi^{\dagger}\chi$, and leading to a non-vanishing correction to the metric in the space $y_I=(\rho, \rho a_I)$.
%The bare mass $m_H$ will be identified with the kaon mass (modulo a string correction), in the discussion of the hyperon spectrum.

The constraint field contribution ${\cal L}_{\Phi_0}[m_H]$ in the last line was analyzed in Appendices A.3 and A.4 of Ref.~\cite{Liu:2021tpq}, and is given by
\begin{align}\label{eq:phi0finalX}
{\cal L}_{\Phi_0}[m_H]= -\frac{1}{8}J_0^\dagger\frac{1}{-D_M^2+m_H^2} J_0\ ,
\end{align}
with the {\it non-local} source $J_0$  given in (\ref{J0X}).
It follows from  the Gauss constraint on the flavor gauge field in bulk,
and is by far the most involved to unravel. For convenience, we detail its analysis in Appendix~\ref{app_phi0},\ref{app_expansion}.
%and will be given explicitly in the two extreme limits of $m_H=0$ and $m_H\rightarrow\infinity}$.
Aside from the explicit mass dependence in (\ref{FULL}), there is an implicit mass dependence in ${\cal L}_{\Phi_0}[m_H]$
which we have noted in the argument. Since the strange mass is intermediate between light $u,d$ and heavy $c,b$,
we will address the implicit mass dependence  in ${\cal L}_{\Phi_0}[m_H]$ both in the light $m_H\rightarrow 0$, and heavy
$m_H\rightarrow \infty$.

The Coulomb contribution ${\cal L}_{\rm Coulomb}$  was originally detailed in Appendix~B in~\cite{Liu:2021tpq},
and for convenience, briefly reviewed in Appendix~\ref{app_coulomb}, with the result

\begin{align}\label{eq:coulombfinalX}
{\cal L}_{\rm Coulomb}=-J^\dagger_C\frac{1}{2\left(-aN_c \nabla^2+f^2\chi^{\dagger}\chi\right)}J_C \ ,
\end{align}
The {\it non-local} source $J_C=\left(\rho^{cl}+\rho\right)$ is given in (\ref{RHOCL}-\ref{RHO}).
Throughout, the Coulomb contribution which is small, will be mostly ignored.
It is a correction to be added  in perturbation theory to the modular
Hamiltonian, and assessed only at the end.

The holographic heavy kaon mass in the large mass limit, is given by~\cite{Liu:2016iqo}
\begin{align}
\label{KKX}
M_K=m_H+\frac{M_{\rm KK}}{2\sqrt{2}} \ ,
\end{align}
with $m_H$ the bare mass of the kaon doublet, and the Kaluza-Klein scale $M_{KK}=475$ MeV. In what
will follow, $m_H\sim M_K$, unless specified otherwise.
%The final numerical results will use (\ref{KKX}).

Note that a naive expansion of the Coulomb and Gauss constraint contributions in (\ref{FULL})
as shown in  Appendix~\ref{app_naive},   leads to a degenerate but stable hyperon spectrum to order $m_H^0$, but  unstable
at sub-leading order. The unexpanded constraints produce a stable hyperon spectrum as we detail below.

\subsection{ ${\cal L}_{\Phi_0}[0]$ and no Coulomb}

We start the analysis of (\ref{FULL}) by considering the simple case  with $m_H=0$ {\it only} in the Gauss constraint  or ${\cal L}_{\Phi_0}[0]$,
and no Coulomb back-reaction.  Both approximations will be revisited below.  With this in mind, the modular Lagrangian simplifies

\bea
\label{QUAD1}
{\cal L}_{\rm qua}=&&\frac{1}{2}\dot \chi^{\dagger}\dot \chi+\frac{3i}{\tilde \rho^2}\chi^{\dagger}\dot \chi-\frac{37+12\frac{Z^2}{\rho^2}}{192}\chi^{\dagger}\chi\nonumber\\
&&+\frac{99i}{40\tilde \rho^2}\chi^{\dagger}\tau^a\chi \chi^a-\frac{75}{8\tilde \rho^4}\chi^{\dagger}\chi \ .
\eea
Note that without the spin-orbit coupling, we have

\begin{align}
\label{SP0}
{\cal L}_0=\frac{1}{2}\dot \chi^{\dagger}\dot \chi+\frac{3i}{\tilde \rho^2}\chi^{\dagger}\dot \chi-\frac{37+12\frac{Z^2}{\rho^2}}{192}\chi^{\dagger}\chi-\frac{75}{8\tilde \rho^4}\chi^{\dagger}\chi \ .
\end{align}
By setting the kaon modular variable  $\chi$ as

\begin{align}
\label{chi12}
   \chi=\left( \begin{array}{c}
       x_1+iy_1    \\
        x_2+iy_2
    \end{array}\right) \ ,
\end{align}
(\ref{SP0}) can be  written as two harmonic oscillators coupled to magnetic field

\bea
\label{QUAD2}
    {\cal L}_0=&&\frac{1}{2}(\dot {\vec{x_1}}^2+\dot {\vec{x_2}}^2)+\omega_c (y_1\dot x_1-x_1\dot y_1+y_2\dot x_2-x_2\dot y_2)\nonumber\\
    &&-\frac{m_H^2+\Omega^2}{2}(\vec{x_1}^2+\vec{x_2}^2) \ ,
\eea
where we have defined
\begin{align}
\Omega^2=\frac{75}{4\tilde \rho^4}+\frac{37+6\sqrt{6}\frac{1}{\tilde \rho^2}}{96} \ \  , \ \   \omega_c=\frac{3}{\tilde \rho^2} \ .
\end{align}
This observation will be exploited next.

\section{Hyperon spectrum}~\label{sec_spectrum1}

Following on the preceding arguments, we now analyze the modular Hamiltonian stemming from
(\ref{QUAD1}). Without the spin-orbit contributions as we noted in (\ref{QUAD2}), it maps on the well-known Landau
problem in 2-dimensions.  In this regime, the hyperons are stable but degenerate.
The spin-orbit contribution modifies  the  potential in the holographic $\rho$-direction, and lifts the hyperon degeneracy.

\subsection{Landau problem}

For the modular Lagrangian (\ref{QUAD2}), the pertinent Schroedinger equation reads
\begin{align}
H\phi_n(x)=E\phi_n \ ,
\end{align}
with the modular Hamiltonian
\begin{align}
H=\frac{1}{2}D_i^{\dagger}D_i+\frac{\omega^2}{2}\bar z z \ ,
\end{align}
with $z=x+iy$. The long derivative is
$D_i=\partial_i-iA_i$, with the U(1) gauge field $A_i=\omega_c(y,-x)$. We now define the operators
\bea
a&=&\frac{i}{2\sqrt{\omega_c}}(D_x-iD_y)\nonumber\\
&=&\frac{i}{2\sqrt{\omega_c}}(\partial_x-i\partial_y+\omega_c(x-iy)) \ , \nonumber\\
b&=&\frac{-i}{2\sqrt{\omega_c}}(-\partial_x-i\partial_y- \omega_c(x+iy)) \ ,
\eea
which diagonalizes the kinetic contribution
\begin{align}
\frac{1}{2}D_i^{\dagger}D_i=\omega_c (2a^{\dagger}a+1) \ ,
\end{align}
For the harmonic contribution, we note that
\begin{align}
b^{\dagger}-a=-i\sqrt{\omega_c}(x-iy) \ ,
\end{align}
hence the Hamiltonian can be written as
\begin{align}
\label{HX1}
H=\omega_c (2a^{\dagger}a+1)+\frac{\omega^2}{2\omega_c}(b^{\dagger}-a)(b-a^{\dagger}) \ .
\end{align}
The Hamiltonian (\ref{HX1}) can then be diagonalized with the help of the following Bogoliubov transformation
\begin{align}
a^{\dagger}=\cosh \theta A^{\dagger}+\sinh \theta B \ , \\
b^{\dagger}= \cosh \theta B^{\dagger}+\sinh \theta A \ .
\end{align}
Using $[A,A^{\dagger}]=[B,B^{\dagger}]=1$ and $[A,B]=[A,B^{\dagger}]=0$,  which preserves the commutation relations,
we can fix the value of  $\theta$  as
\begin{align}
\tanh 2\theta=\frac{2\alpha}{1+2\alpha}\ , \,\,\, \alpha=\frac{\omega^2}{4\omega_c^2} \ .
\end{align}
The modular Hamiltonian  without spin-orbit coupling is then diagonalized as
\begin{align}
\label{HX2}
H=\frac{\Omega_++\Omega_-}{2} + \Omega_+A^{\dagger}A+\Omega_{-}B^{\dagger}B \ ,
\end{align}
with
\begin{align}
\Omega_{\pm}=\sqrt{m_H^2+\Omega^2 +\omega_c^2}\pm \omega_c \  .
\end{align}
In the next subsection we will explore the spin-orbital contribution.

\subsection{Spin-orbit}

For fixed modular variable $\tilde\rho$, the holographic spectrum  without spin-orbit following from (\ref{HX2})
is harmonic. Since the modular coordinate $\chi$ is quantized as a boson, the net spin and isospin of the
hyperon core is determined by the instanton quantum moduli with $[IJ^P]=[\frac12\frac 12^+]$ assignment,
in the absence of spin-orbit effects. With
spin-orbit contributions, the resulting hyperon states carry $[\frac 12\oplus \frac 12,\big( \frac 12\oplus l\big)^+]$
assignments, for even $l$. We now proceed to analyze the dynamical effects of the spin-spin and spin-orbit effects.

\subsubsection{The $l=0$ state}

For $l=0$ and $J=\frac 12$, the energy level with $n$ $B^{\dagger}$ excitations is
\begin{align}
 E_n-E_0=n\left(\sqrt{m_H^2+\frac{111}{4\tilde \rho^4}+\frac{37+12\frac{ Z^2}{\rho^2}}{96}}-\frac{3}{\tilde \rho^2}\right) \ ,
\end{align}
and the lowest one is archived for $n=1$. To proceed we need to fix the  $\rho$ wave function. For that, the induced potential is given by $\Omega_-$
to which we add   the harmonic oscillator potential term $\frac{1}{2}\omega_\rho^2\tilde \rho^2$,
plus  the quartic term $-\frac{3i}{5\tilde \rho^4 }\chi^{\dagger}\tau^a\dot \chi \chi^{\dagger}\tau^a \chi$  as in  Eq.~43 in~\cite{Liu:2021tpq}. The result is
\begin{widetext}
\begin{align}
\label{POTX1}
V(\tilde \rho)=\frac{1}{2}\omega_\rho^2\tilde \rho^2+\sqrt{m_H^2+\frac{111}{4\tilde \rho^4}+\frac{37+12\frac{Z^2}{\rho^2}}{96}}-\frac{3}{\tilde \rho^2}+\frac{9(\sqrt{m_H^2+\frac{111}{4\tilde \rho^4}+\frac{37+12\frac{Z^2}{ \rho^2}}{96}}-\frac{3}{\tilde \rho^2})}{5(m_H^2+\frac{111}{4\tilde \rho^4}+\frac{37+12\frac{\tilde Z^2}{\tilde \rho^2}}{96})\tilde \rho^4} \ .
\end{align}
\end{widetext}
We note that (\ref{POTX1})  is stable for small $\rho$. The additional parameter
$\delta$ captures a spin-spin ordering ambiguity to be discussed below. With this in mind,
%The lats term is due to the $(\chi^{\dagger}\tau^a\chi)^2$ contribution, which we will discuss below.
and using the estimate
\begin{align}
\frac{Z^2}{\rho^2} \approx \sqrt{\frac{3}{2}}\frac{1}{\tilde \rho^2}
\end{align}
the splitting between $\Lambda^0$ and nucleon,  can be solved numerically for $\delta=1$. The result is
\begin{align}
M_{\Lambda^0}-M_{N}=0.237M_{\rm KK} \ ,
\end{align}
For $M_{\rm KK}=0.475$ GeV,  the splitting is about  $112.7$ MeV, smaller than the empirical
splitting of  $177$ MeV.  This is reasonable, since the omitted Coulomb back-reaction is positive
(see below).

\subsubsection{The $l=2$ state}
For the  $l\ne 0$ cases,  the quantization needs to be considered more carefully, as
operator ordering issues arise. Indeed, we note that the spin operator in the Bogoliubov
transformed basis, reads

\begin{align}
\chi^{\dagger}\tau^a \chi =\frac{1}{\sqrt{m_H^2+\Omega^2+\omega_c^2}}(A^{\dagger}-B)_i\tau^a_{ij} (A-B^{\dagger})_j \ .
\end{align}
The spin-spin and spin-orbit effects will be treated in first order perturbation theory.
 When evaluating the average of $\chi^{\dagger}\tau^a \chi$, one recovers the standard Schwinger representation of a $\frac{1}{2}$-spin
\begin{align}
S^a=\frac{1}{2}B^{\dagger}_i\tau^a_{ij} B_j \ ,
\end{align}
with  $A_1, A_2$  constructed using $(x_1,y_1)$ and $(x_2, y_2)$, respectively.
When evaluating $(\chi^{\dagger}\tau^a\chi)^2$, without normal ordering, gives
\bea
&&\langle 0|B_i (A^{\dagger}-B)\tau^a (B^{\dagger}-A^{\dagger}) \nonumber\\
&&\times(A^{\dagger}-B)\tau^a (B^{\dagger}-A) B_i^{\dagger}|0\rangle =12 \ ,
\eea
for $i=1,2$. With normal ordering, the result is different
\bea
&&\langle 0|B_i :(A^{\dagger}-B)\tau^a (B^{\dagger}-A^{\dagger})\nonumber\\
&&\times  (A^{\dagger}-B)\tau^a (B^{\dagger}-A): B_i^{\dagger}|0\rangle =6  \ .
\eea
The  normal ordering ambiguity  is captured by a c-number  $\delta$.  The third and perhaps most physical
choice,  amounts to dropping the anti-particle contribution to the spin  through  $A,A^\dagger$,
\begin{align}
\langle 0|B_i  B\tau^aB^{\dagger} B\tau^a B^{\dagger} B_i^{\dagger}|0\rangle\nonumber =3 \ .
\end{align}
which is ordering free. This choice corresponds to $\delta=1$, and will be subsumed throughout.

%Using these approximation, one has the spin part of the Hamiltonian.

For $l=2,4,..$, one has $J=(l\pm 1)/2$. We first consider the $J=(l-1)/2$ case. Following our recent arguments in~\cite{Liu:2021tpq} (e.g. Eqs. 44-45),
the effective potential reads

\begin{widetext}
\bea
    && V\bigg(J=\frac{l-1}{2},\tilde \rho\bigg)=\frac{1}{2\tilde \rho^2}\left(l(l+2)-\frac{(l+2)\alpha N_c}{\sqrt{m_H^2+\Omega^2+\omega_c^2}\tilde \rho^2}+\frac{3 \alpha^2N_c^2}{4(m_H^2+\Omega^2+\omega_c^2)\tilde \rho^4}\right)\nonumber \\
    &+&\frac{\omega_\rho^2}{2}\tilde \rho^2+\sqrt{m_H^2+\Omega^2+\omega_c^2}-\omega_c +\frac{9(\sqrt{m_H^2+\frac{111}{4\tilde \rho^4}+\frac{37+12\frac{Z^2}{ \rho^2}}{96}}-\frac{3}{\tilde \rho^2})}{5(m_H^2+\frac{111}{4\tilde \rho^4}+\frac{37+12\frac{ Z^2}{\rho^2}}{96})\tilde \rho^4} \ ,
\eea
\end{widetext}
with $\alpha=\frac{33}{10}$. The ${1}/{m_H^2}$ term due to the spin-orbit coupling is kept to maintain stability at small $\rho$.
The change of the potential as one increases $m_H$ tends to decrease for larger $l$. For $l=2$, the
the  potentials at $m_H=2$ and $m_H=\infty$ differ moderately, but the net difference is small.

Similarly, in the $J=\frac{l+1}{2}$ case the effective potential is

\begin{widetext}
\bea
&&V\bigg(J=\frac{l+1}{2},\tilde \rho\bigg)=\frac{1}{2\tilde \rho^2}\left(l(l+2)+\frac{l\alpha N_c}{\sqrt{m_H^2+\Omega^2+\omega_c^2}\tilde \rho^2}+\frac{3 \alpha^2N_c^2}{4(m_H^2+\Omega^2+\omega_c^2)\tilde \rho^4}\right)\nonumber \\
    &&+\frac{\omega_\rho^2}{2}\tilde \rho^2+\sqrt{m_H^2+\Omega^2+\omega_c^2}-\omega_c+ \frac{9(\sqrt{m_H^2+\frac{111}{4\tilde \rho^4}+\frac{37+12\frac{ Z^2}{ \rho^2}}{96}}-\frac{3}{\tilde \rho^2})}{5(m_H^2+\frac{111}{4\tilde \rho^4}+\frac{37+12\frac{Z^2}{\rho^2}}{96})\tilde \rho^4} \  .
\eea
\end{widetext}
For $\delta=1$ and  $m_H=M_{KK}$, a numerical analysis for the hyperon states gives

\bea
&&[J=\frac{1}{2}, l=2, I=1]: M(\Sigma_{s}(1\frac{1}{2}^+))-M_N=302{\text MeV} \nonumber \\
&&[J=\frac 32, l=2, I=1]:  M(\Sigma_{s}(1\frac{3}{2}^+))-M_N=501{\text MeV} \nonumber\\
\eea
which are to be compared to  the measured values of
 $254$ MeV and $444$ MeV. The  splitting between the centroid is much more accurate
\begin{align}
M(\Sigma_{s}(1\frac{3}{2}^+))-M(\Sigma_{s}(1\frac{1}{2}^+))=199 {\text MeV} \ ,
\end{align}
compared to  $191$ MeV, empirically.

\begin{widetext}
\section{Hyperon spectrum revisited}~\label{sec_spectrum2}

%{The spectrum for $m_H\rightarrow \infty$ approach}
We now consider the hyperon spectrum with spin-orbit effect, but with
${\cal L}_{\Phi_0}[m_H]$ in the opposite limit of large $m_H$ for comparison.
The details of ${\cal L}_{\Phi_0}[m_H]$ are presented in Appendix~\ref{app_phi0},
including its closed form results in the heavy mass limit.

\subsection{Without Coulomb}

 In this case the potentials in the
holographic $\rho$-direction are modified as follows

\bea
V_{l=0}(\tilde \rho)=&&\frac{1}{2}\omega_\rho^2\tilde \rho^2+\bigg(m_H^2+\frac{9}{\tilde \rho^4}(1+\frac{4.11}{m_H^2\tilde \rho^2})+\frac{37+12\frac{ Z^2}{ \rho^2}}{96}\bigg)^{\frac 12}-\frac{3}{\tilde \rho^2}\nonumber\\
&&+\frac{9\bigg(\bigg({m_H^2+\frac{9}{\tilde \rho^4}(1+\frac{4.11}{m_H^2\tilde \rho^2})+\frac{37+12\frac{ Z^2}{ \rho^2}}{96}}\bigg)^{\frac 12}-\frac{3}{\tilde \rho^2}\bigg)}{5(m_H^2+\frac{9}{\tilde \rho^4}(1+\frac{4.11}{m_H^2\tilde \rho^2})+\frac{37+12\frac{ Z^2}{\rho^2}}{96})\tilde \rho^4} \ .
\eea
for $l=0$, and for $l=2$
\bea
&&V_{l}(J=\frac{l-1}{2},\tilde \rho)=V_{l=0}(\tilde \rho)+\frac{1}{2(1+\frac{1}{2m_H\tilde \rho^2})\tilde \rho^2}\left(l(l+2)-\frac{(l+2)\tilde \alpha N_c}{\sqrt{m_H^2+\tilde \Omega^2+\omega_c^2}\tilde \rho^2}+\frac{3\tilde \alpha^2N_c^2}{4(m_H^2+\tilde \Omega^2+\omega_c^2)\tilde \rho^4}\right) \nonumber\\
%\end{align}
%and
%\begin{align}
&&V_{l}(J=\frac{l+1}{2},\tilde \rho)=V_{l=0}(\tilde \rho)+\frac{1}{2(1+\frac{1}{2m_H\tilde \rho^2})\tilde \rho^2}\left(l(l+2)+\frac{(l)\tilde \alpha N_c}{\sqrt{m_H^2+\tilde \Omega^2+\omega_c^2}\tilde \rho^2}+\frac{3 \tilde \alpha^2N_c^2}{4(m_H^2+\tilde \Omega^2+\omega_c^2)\tilde \rho^4}\right) \nonumber\\
\eea

%\begin{widetext}
\begin{table}[h]
\caption{Hyperon and exotic spectrum}
\begin{center}
\begin{tabular}{cccccccccc}
\hline
\hline
$B$ & $IJ^P$  &  $l$  & $n_\rho$ & $n_z$  & Mass(small)& Mass(small with Coulomb)& Mass(large)&Mass(large with Coulomb) & Exp-MeV \\
\hline
\hline
$\Lambda_s$       &$0{\frac 12}^+$ & 0  & 0&  0& 962  & 1182  & 974  & 1152  & 1115  \\
$\Sigma_s$        &$1{\frac 12}^+$ & 2  & 0&  0& 1134 & 1315  & 1149 & 1306  & 1192  \\
                  &$1{\frac 32}^+$ & 2  & 0&  0& 1346 & 1472  & 1254 & 1398  & 1387  \\
$\Theta_s^+$      &$0{\frac 12}^+$ & 0  & 0&  0&  -   & 1617  & -    & 1599  &   \\
\hline
\hline
\end{tabular}
\end{center}
%\caption{Hyperon spectrum in the small and large $m_H=0.68 M_{KK}$ approximation, without and with the Coulomb contribution. See text.}
\label{tab_bindtetb}
\end{table}%
%\end{widetext}

\end{widetext}
with
\bea
\tilde \alpha=&&\frac{13}{10}+\frac{162}{35\tilde m_H^2 \tilde \rho^4 } \ , \nonumber\\
\tilde \Omega^2=&&\frac{37+6\sqrt{6}\frac{1}{\tilde \rho^2}}{96}+\frac{9*4.11}{m_H^2\tilde \rho^2} \ .
\eea
Also, there is a modification to the curvature in the $\rho$ direction. We should also include the leading warping contribution at large $m_H$, the details of which are identical to those presented in~\cite{Liu:2021tpq}. With this in mind, the hyperon spectrum is now given by

\bea
&&[J=\frac{1}{2}, l=0, I=0]:  M(\Lambda)-M_N=68.1{\text MeV} \ ,\nonumber \\
&&[J=\frac{1}{2}, l=2,  I=1]: M(\Sigma_{s}(1\frac{1}{2}^+))-M_N=289{\text MeV}  \nonumber\\
&&[J=\frac{3}{2}, l=2 , I=1]:   M(\Sigma_{s}(1\frac{3}{2}^+))-M_N=400{\text MeV} \nonumber\\
\eea
The $J=\frac{1}{2}$ $\Sigma$ state is pushed up,  and the  $J=\frac{3}{2}$ $\Sigma$ state is pushed down, with a
split in the centroid
\begin{align}
\frac{M_{\Sigma}(1\frac{1}{2}^+)+M_{\Sigma}(1\frac{3}{2}^+)}{2}-M_N=344{\text MeV} \ ,
\end{align}
which is close to the empirical  value of 349 MeV.

\subsection{With Coulomb}

As we indicated earlier, throughout we assumed $m_H\sim M_{KK}$ in (\ref{KKX}). Here we correct for
this shortcoming, with

\begin{align}
m_H=0.68 M_{\rm KK} \ .
\end{align}
and  $M_{\rm KK}=475$ MeV fixed by the light baryon spectrum~\cite{Hashimoto:2009st}.

Also, the neglected Coulomb contribution can be estimated in perturbation
theory,  and in the heavy meson mass  limit, it is about

\begin{align}
V_{\rm C}\approx \frac{83}{30 \tilde \rho^2} \ .
\end{align}
This provides for an upper bound estimate.

With in mind, the modified hyperon masses are

\bea
&&[J=\frac{1}{2},  l=0, I=0]:  M(\Lambda)-M_N=214{\text MeV} \ , \nonumber\\
&&[J=\frac{1}{2},l=2,   I=1]:  M(\Sigma_{s}(1\frac{1}{2}^+))-M_N=368{\text MeV}\nonumber \\
&&[J=\frac{3}{2}, l=2 ,  I=1]: M(\Sigma_{s}(1\frac{3}{2}^+))-M_N=460{\text MeV}\nonumber\\
\eea
The experimental values are $177$ MeV, $254$ MeV and $440$ MeV respectively,  with
$37$ MeV, $133$ MeV and $20$ MeV differences.

Using the corrected value of $m_H$ above,
and the upper estimate for  the Coulomb correction, in Table~\ref{tab_bindtetb}, we collect  all hyperon masses for the three approximations  presented earlier.
The chief observation is that the large mass analysis without Coulomb corrections appear closer to the empirical
values of the lowest three empirical hyperons, without any adjustable parameter. These results are to be compared
 to those reported by Callan and Klebanov using the Skyrme model~\cite{Callan:1985hy, Callan:1987xt}, with also no Coulomb corrections.

We recall that in the present holographic construction,  the relative splitting between the hyperons, and   the splitting
of the hyperon centroid from the nucleon, which eliminate much of the uncertainty in $M_{KK}$, are in remarkable
agreement with the empirically reported splittings.

\subsection{Exotics}

 This approach extends to light multiquark exotics with open or hidden strangeness, much like the heavier multiquark
 exotics with open or hidden charm and bottom discussed in~\cite{Liu:2019mxw,Liu:2019yye,Liu:2021tpq,Liu:2021ixf}.  In particular,
an  estimate of the mass of  the strange pentaquark $\Theta_s^+$ (the exotic $uudd\bar s$),
is given in Table~\ref{tab_bindtetb}. The mass of about 1600 MeV, stems mainly from the $\Omega_+$ frequency (anti-particle)
 which is $\frac{6}{\tilde \rho^2}$ higher than the $\Omega_-$ (particle). (Recall that
 the effective magnetic field induced by the bulk Chern-Simons interaction,  is repulsive for particles,
 and attractive for anti-particles).  An additional repulsion of about $\frac{3}{\tilde \rho^2}$  stems from
 the Coulomb back-reaction in the heavy mass estimate. A $\Theta_s^+$ of about
 1600 MeV lies above the $nK^+$ threshold of 1434 MeV, and is  unstable against strong decay.
 This result is consistent with the fact that the proposed $\Theta_s^+$ state~\cite{Jaffe:1976ii,Praszalowicz:1987em,Diakonov:1997mm,Praszalowicz:2003ik},
 is in so far unaccounted for experimentally.

%We should also mention that there are two adjustable parameters in our current approach, $\delta$ and $m_{\rm KK}$. We have found that for $\delta=1$, the $M_{\rm KK}$ around $475$ MeV produces the best spectrum. While it is possible to increase $\delta$ to have a better result for $\Lambda$ in the approaches neglects the Coulomb back-reaction, it will tend to push the $\Sigma$ states higher.

\section{Conclusions}~\label{sec_conclusions}

In the holographic construction presented in~\cite{Liu:2016iqo,Liu:2017xzo,Liu:2017frj}, heavy hadrons are described in bulk
using a set of degenerate $N_f$ light D8-D$\bar 8$ branes  plus one heavy probe brane in the
cigar-shaped geometry that spontaneously breaks chiral symmetry. This construction enforces both chiral and
heavy-quark symmetry and describes well  the low-lying
heavy-light mesons, baryons and multi-particle exotics~\cite{Liu:2019mxw,Liu:2019yye,Liu:2021tpq,Liu:2021ixf}. Heavy  hadrons
whether standard or exotics,  are composed of heavy-light mesons bound to a core instanton in bulk.

In~\cite{Liu:2017frj} the analysis of the  hyperon spectrum was carried to order $m_H^0$
where  spin effects are absent. In this analysis, we have now carried the analysis at next to
leading order in $1/m_H$ where the spin-orbit and spin corrections are manifest. In contrast
to~\cite{Liu:2017frj}, the modular fields were quantized as bosons and not fermions. The
quantized Hamiltonian describes a particle in an external 2-dimensional magnetic field, with
spin-orbit coupling.

The hyperon spectrum with the Gauss constraint treated in both the heavy and light kaon mass
limit, shows very small changes. It is in overall agreement with the empirical hyperon spectrum,
and is  much improved in comparison to the analysis in~\cite{Liu:2017frj}, where the strange mass
was  analysed perturbatively.
This construction allows for the description of  multiquark exotics with strangeness, and shows that
the contentious exotic $\Theta_s^+$ is unbound. In  a way, this construction should be regarded as the dual
of the  improved Callan-Klebanov construction for hyperons, as bound kaon-Skyrmions~\cite{Callan:1985hy,Callan:1987xt}.

%By refining the Kaluza-Klein scale $M_{KK}$ from 1 GeV used in~\cite{Liu:2017xzo,Liu:2017frj}  to 0.475 GeV used here,
%a rich spectrum with single- and double-heavy baryons emerges with fair  agreement with the empirically observed
%states, including the newly reported charm pentaquark states by LHCb.

\vskip 1cm
{\bf Acknowledgements}

This work is supported by the Office of Science, U.S. Department of Energy under Contract No. DE-FG-88ER40388,
by the Polish National Science Centre (NCN) Grant UMO-2017/27/B/ST2/01139, and by the 2021-N17/MNS/35 grant of the Faculty of Physics, Astronomy and Applied Computer Science of the Jagiellonian University.

\appendix

\section{\\Derivation of ${\cal L}_{\Phi_0}[m_H]$}~\label{app_phi0}

%In this Appendix,  we review the constraint field contribution.
%More details can be found in  Appendices A.3 and A.4 in~\cite{Liu:2021tpq}.

For a generic kaon mass of order $m_H$, we must include
its  contribution in the Gauss law constraint as captured by the
time component $\Phi_{M=0}$ of the heavy-light vector field.
This is the most difficult term to unravel to order $1/m_H$.
  For that, we recall from Appendices A.3 in~\cite{Liu:2021tpq},
 that  the constraint equation  for $\Phi_0$ is

\bea
\label{AP10}
&&(-D_M^2+m_H^2) \Phi_0+2F_{M0}\Phi_M\nonumber\\
&&-\frac{i}{16\pi^2 a}F_{PQ}(\partial_P+A_P) \Phi_Q=0 \ .
\eea
after using the self-dual condition for $F$. Using  the standard relations for $\bar \sigma_{MN}$, we  have for the last two contributions
in (\ref{AP10})

\begin{align}
\label{AP11}
F_{PQ}\partial_P \Phi_Q=\frac{6\rho^2}{(X^2+\rho^2)^2} \frac{1}{r}\frac{ df}{dr}\bar\sigma \cdot X \chi \ ,\nonumber\\
F_{PQ}A_P\Phi_Q=-\frac{6 \rho^2}{(X^2+\rho^2)^3}f\bar\sigma \cdot X\chi  \ .
\end{align}
For the first contribution in (\ref{AP10}) we have

\bea
\label{AP12}
F_{M0}\Phi_M=&&\frac{6f}{(X^2+\rho^2)^2}\left(\rho^2 \bar \sigma \cdot \dot X+\bar \sigma \cdot X \rho\dot\rho\right)\chi\nonumber\\
&&+\chi^aD_M\Phi^a \bar \sigma_M \chi f \ .
\eea
with

\begin{align}
\Phi^a=\frac{1}{2(X^2+\rho^2)}\bar \sigma \cdot X \tau^a \sigma \cdot X \ ,
\end{align}
or more explicitly

\begin{align}
\label{AP11}
\chi^aD_M\Phi^a \bar \sigma_M \chi f=  \frac{3\rho^2 f}{(X^2+\rho^2)^2}\bar \sigma \cdot X \tau^a \chi \chi^a \ .
\end{align}
Inserting (\ref{AP11}-\ref{AP12}) into (\ref{AP10})  we  have

\begin{align}
\label{AP12}
(-D_M^2 +m_H^2)\Phi_0+J_0=0 \ ,
\end{align}
with

\bea
J_0=&&\frac{12f}{(X^2+\rho^2)^2}\left(\rho^2 \bar \sigma \cdot \dot X+\bar \sigma \cdot X \rho\dot\rho\right)\chi\nonumber\\
&&+ \frac{6f \rho^2}{(X^2+\rho^2)^2}\bar \sigma \cdot X \tau^a \chi \chi^a \nonumber \\
&&+\frac{3i}{2\pi^2 a}\frac{\rho^2 f}{(X^2+\rho^2)^3} \bar \sigma \cdot X \chi + \frac{2f}{r}\frac{\partial \hat A_0}{\partial r} \bar \sigma \cdot X \chi\nonumber\\
\eea
 the source for $\Phi_0$
 \begin{align}
 {\cal L}_{\Phi_0}= \frac{1}{8}\int d^4X J_0^{\dagger}(X)\Phi_0(X) \ .
\end{align}
 In this equation the Abelian part of $F_{N0}$ has been included.  Since
\begin{align}
    \frac{1}{r}\frac{\partial \hat A_0}{\partial r}=\frac{i}{4\pi^2 a}\frac{1}{(X^2+\rho^2)^2}\left(1+\frac{2\rho^2}{X^2+\rho^2}\right)
\end{align}
one finally has

\bea
\label{J0X}
   J_0=&&\frac{12f}{(X^2+\rho^2)^2}\left(\rho^2 \bar \sigma \cdot \dot X+\bar \sigma \cdot X \rho\dot\rho\right)\chi\nonumber\\
   &&+ \frac{6\rho^2f}{(X^2+\rho^2)^2}\bar \sigma \cdot X \tau^a \chi \chi^a \nonumber \\
&&+\frac{i}{2\pi^2 a}\frac{f}{(X^2+\rho^2)^2}\left(1+\frac{5\rho^2}{X^2+\rho^2}\right) \bar \sigma \cdot X \chi\nonumber\\
\eea

%After solving for  $\Phi_0$, the contribution to the modular Lagrangian reads

To solve (\ref{AP12}),  we need the massive spin-0 propagator in the instanton background

\bea
\label{G2XY}
G_{2}(X,Y)=\langle X|\frac{1}{-D_M^2+m_H^2}|Y\rangle \ ,
%\nonumber\\
%&&\Phi_0(X)=-\int d^4 Y G_{m_H}(X,Y)J_0(Y) \ ,
\eea
in terms of which the Gauss law constraint yields the modular Lagrangian contribution (\ref{AP12}) in the form

\begin{align}\label{eq:phi0final}
{\cal L}_{\Phi_0}= -\frac{1}{8}\int d^4X d^4Y J_0^{\dagger}(X)\langle X|\frac{1}{-D_M^2+m_H^2}|Y\rangle J_0(Y) \ ,
\end{align}

\section{\\Expansion of ${\cal L}_{\Phi0}[m_H]$}~\label{app_expansion}

The spin-0 Greens function (\ref{G2XY}) is not known for arbitrary $m_H$,
except for $m_H=0$. Here, we provide a general expression for the different
modular contributions in ${\cal L}_{\Phi0}[m_H]$, and then specialize to the
two extreme cases  of $m_H=0$ and large $m_H$, for which analytical expressions
can be obtained. More specifically, we have

\bea
{\cal L}_{\Phi_0}=&&-\chi^{\dagger}\chi \bigg( \frac{\alpha}{\rho^2}\dot X^2+\frac{\beta}{\rho^2}(\dot \rho^2+\rho^2 a_I^2) \bigg)\nonumber\\
&&+\frac{i\gamma}{\tilde \rho^2}\chi^{\dagger}\tau^a\chi \chi^a-\frac{\delta}{\tilde \rho^4}\chi^{\dagger}\chi \ ,
\eea
with the coefficients
\bea
\alpha&=&\frac{12^2}{4}\int d^4X d^4Y \frac{f(X)g_1(X,Y) f(Y)}{(X^2+1)^2(Y^2+1)^2} \ , \nonumber\\
\beta&=&\frac{12^2}{16} \int d^4X d^4Y \frac{f(X) g_2(X,Y) f(Y)}{(X^2+1)^2(Y^2+1)^2} \ , \nonumber\\
\gamma&=&\frac{48N_c}{8}\int d^4X d^4Y  \frac{f(X)}{(X^2+1)^2}\nonumber\\
&&\times\left(1+\frac{5}{X^2+1}\right)\frac{g_2(X,Y)f(Y)}{(Y^2+1)^2}  \ ,\nonumber \\
\delta&=&\frac{64N_c^2}{16}\int d^4X d^4Y \frac{f(X)}{(X^2+1)^2}\nonumber\\
&&\times \left(1+\frac{5}{X^2+1}\right)g_2(X,Y)\nonumber\\
&&\times \frac{f(Y)}{(Y^2+1)^2}\left(1+\frac{5}{Y^2+1}\right)\ .
\eea
Here the scalar functions trace over the spin-0 propagator
 \bea
 g_1(X,Y)&=& {\rm tr} G_2(X,Y) \ , \nonumber\\
 g_2(X,Y)&=&{\rm tr}\left(\sigma \cdot X G_2(X,Y)\bar \sigma \cdot Y \right) \ ,
 \eea
after the re-scaling  $\rho \rightarrow 1$ and $m_H \rightarrow m_H \rho^2$.
For $m_H=0$, the expressions will be quoted explicitly below. For $m_H$ large, spin-0 propagator
is zero-mode free, and can be approximated by its free part
 \begin{align}
 G_2(X,Y)\rightarrow \int \frac{d^4k}{(2\pi)^4}\frac{e^{ik\cdot (X-Y)}}{k^2+m_H^2} \ ,
 \end{align}
Using  the Fourier transforms

\bea
&&\frac{f(X)}{(X^2+1)^2}=\int\frac{d^4k}{(2\pi)^4}g_1(k)e^{ik\cdot X} \ , \nonumber\\
&&\frac{f(X)}{(X^2+1)^2}\left(1+\frac{5}{X^2+1}\right)=\int\frac{d^4k}{(2\pi)^4}g_2(k)e^{ik\cdot X} \ , \nonumber\\
\eea
we have
\begin{align}
&g_1(k)=\frac{4\sqrt{2}\pi}{15}e^{-|k|}(1+|k|) \ ,\nonumber \\
&g_2(k)=\frac{4\sqrt{2}\pi}{105}e^{-|k|}(5|k|^2+22|k|+22) \ .
\end{align}
so that
\bea
\alpha&=&\frac{12^2}{2}\int \frac{d^4k}{(2\pi)^4}\frac{|g_1(k)|^2}{k^2+\tilde m_H^2} \ , \nonumber\\
\beta&=&\frac{12^2}{8}\int \frac{d^4k}{(2\pi)^4}\frac{|\nabla g_1(k)|^2}{k^2+\tilde m_H^2} \ , \nonumber\\
\gamma&=&\frac{48N_c}{4}\int \frac{d^4k}{(2\pi)^4}\frac{\nabla g_1(k)\cdot \nabla g_2(k)}{k^2+\tilde m_H^2} \ ,\nonumber \\
\delta&=&\frac{64N_c^2}{8}\int \frac{d^4k}{(2\pi)^4}\frac{|\nabla g_2(k)|^2}{k^2+\tilde m_H^2} \ ,
\eea
\\
\\
{\bf Large $m_H$:}
\\
\bea
\alpha=&&\frac{12^2}{\pi^2 \tilde m_H^2}\int d^4 X \frac{1}{(X^2+1)^7}=\frac{24}{5\tilde m_H^2} \ ,\nonumber \\
\beta=&&\frac{12^2}{4\pi^2 \tilde m_H^2}\int d^4 X \frac{X^2}{(X^2+1)^7}=\frac{6}{5\tilde m_H^2} \ , \nonumber \\
\gamma=&&\frac{48N_c}{2\pi^2 \tilde m_H^2}\int d^4 X \frac{X^2}{(X^2+1)^7}\nonumber \\
&&\times\left(1+\frac{5}{X^2+1}\right)=\frac{54N_c}{35\tilde m_H^2} \ , \nonumber \\
\delta=&&\frac{64N_c^2}{4\pi^2 \tilde m_H^2}\int d^4 X \frac{X^2}{(X^2+1)^7}\nonumber \\
&&\times \left(1+\frac{5}{X^2+1}\right)^2=\frac{146N_c^2}{35\tilde m_H^2} \ .
\eea
This will actually contribute to order $\frac{1}{m_H^3}$ after the re-scaling in $\chi$.
\\
\\
{\bf Zero $m_H$:}
\\
%One should also notice that for $m_H=0$, one has
\bea
&&\alpha=\beta=\frac{1}{4} \ , \nonumber\\
&&\gamma=\frac{N_c}{2} \ , \nonumber\\
&&\delta=\frac{25N_c^2}{24} \ .
\eea

% \section{The ${\cal L}_{\rm Coulomb}$}

\section{\\ Coulomb correction}~\label{app_coulomb}
\label{BACK}

Here we provide a complete treatment of the Coulomb back-reaction contribution (with more details in  Appendix B in~\cite{Liu:2021tpq}).
After re-scaling the U(1) flavor gauge field in bulk
$A_0\rightarrow iA_0$, the Lagrangian for $A^0$ reads
\begin{align}
{\cal L}[A_0]=\frac{aN_c}{2}(\vec{\nabla} A_0)^2+\frac{f^2}{2}\chi^{\dagger}\chi A_0^2+A_0(\rho^{cl}+\rho)
\end{align}
where $\rho^{cl}$ is the classical source (without the modular field $\chi$)
\begin{align}
\label{RHOCL}
\rho^{cl}=aN_c\nabla^2 A^{cl}_0=-\frac{3N_c}{\pi^2}\frac{\rho^4}{(x^2+\rho^2)^4}
\end{align}
and  $\rho$ the quantum source with  the modular field
\begin{align}
\label{RHO}
\rho=&\frac{f^2}{2}i(\chi^{\dagger}\dot \chi-\dot\chi^{\dagger}\chi)+\frac{3}{16\pi^2a}\frac{2\rho^2-X^2}{(X^2+\rho^2)^2} f^2 \chi^{\dagger}\chi \ .
\end{align}
Note that  the contribution
\bea
&&\frac{3}{16\pi^2a}\frac{2\rho^2-X^2}{(X^2+\rho^2)^2}f^2 \chi^{\dagger}\chi \nonumber\\
&&=\frac{3}{16\pi^2a}\frac{f^2\rho^2}{(X^2+\rho^2)^2}\chi^{\dagger}\chi
+\frac{3}{64\pi^2a}\partial_N\left(\frac{x_Nf^2}{(x^2+\rho^2)}\right)\chi^{\dagger}\chi\nonumber\\
\eea
originates solely  from the Chern-Simons term in bulk.

Given the action for $A_0$, at  the minimum we have
\begin{align}\label{eq:coulombfinal}
{\cal L}_{\rm Coulomb}=-J_C\frac{1}{2\left(-aN_c \nabla^2+f^2\chi^{\dagger}\chi\right)}J_C \ ,
\end{align}
with $J_C=\left(\rho^{cl}+\rho\right)$, which is a complicated function of the scalar  $\chi^{\dagger}\chi$.
More importantly, it yields always a positive mass correction.
Note that the  ${f^2}/{m_H}$ term in the denominator plays the role of a screening mass,
 which can be made more manifest  through a coordinate transformation.

For a general analysis of  the Coulomb correction, we need the
Green's function in the background field,
\begin{align}
G_1(X,Y)=\langle X|\frac{1}{-aN_c \nabla^2+f^2\chi^{\dagger}\chi}|Y\rangle \ ,
\end{align}
In the text, we provide an estimate of this  contribution  in perturbation theory,
with the replacement $\chi^\dagger \chi\rightarrow 1$, for a single bound kaon.

\section{\\Naive $1/m_H$ analysis}~\label{app_naive}

%\subsection{Expansion}
In (\ref{FULL}) both the Gauss law constraint and the Coulomb back-reaction are complicated functions
of the  modular coordinate $\chi$ and $m_H$. Naively, a standard quantum analysis would require
expanding them in $1/m_H$. This expansion, leads to an unstable hyperon
spectrum at next-to-leading order, as we now demonstrate. In a way the charge constraint and screening
should not be expanded, to garantee quantum stability.

Consider (\ref{FULL}) with all terms expanded to order to order ${\cal O}({1}/{m_H^2})$
\begin{align}\label{eq:final1}
{\cal L}={\cal L}_{\rm quadratic}+{\cal L}_{\rm int}
\end{align}
where the quadratic part reads
\bea
{\cal L}_{\rm quadratic}=&&i\chi^{\dagger}\dot \chi+\frac{1}{2m_H}\dot \chi^{\dagger}\dot \chi +\frac{9}{2\tilde \rho^2}\chi^{\dagger}\chi+\frac{9}{2m_H\tilde \rho^2}j \nonumber\\&&+\frac{78}{5m_H\tilde \rho^2} i\chi^{\dagger}\tau^a \chi\chi^a \nonumber \\
&&+\frac{102}{5m_H\tilde \rho^4}\chi^{\dagger}\chi-\frac{37+12\frac{Z^2}{\rho^2}}{192m_H}\chi^{\dagger}\chi \ ,
\eea
and the ``high-order contribution'' ${\cal L}_{\rm int}$ reads

\bea
{\cal L}_{\rm int}=&&-\frac{12}{5m_H\tilde \rho^4}\vec{S}^2
-\frac{2}{3\tilde \rho^2}n^2\nonumber\\
&&+\frac{1}{m_H\tilde \rho^4}\bigg(-\frac{56}{5}n^2+\frac{4}{3}n^3-\frac{4}{3}jn\tilde \rho^2\bigg) \nonumber \\
&&+\frac{1}{m_H^2 \tilde \rho^6}\bigg(-\frac{128 n^4}{45}+\frac{376 n^3}{15}-\frac{4017 n^2}{70}\nonumber\\
&&-jn\tilde \rho^2 \left(\frac{56}{5}-\frac{8}{3}n\right)-\frac{2}{3}j^2 \tilde \rho^4 \bigg)   \ ,
\eea
with
\begin{align}
j=\frac{i}{2}\left(\chi^{\dagger}\dot\chi-\dot\chi^{\dagger}\chi\right),\  n=\chi^{\dagger}\chi \  ,
\end{align}
%We now study the quadratic part.

%\subsection{Quantization}
We now focus on the quadratic part, by
replacing $\chi \rightarrow e^{im_Ht}\sqrt{m_H}$,  so that
\bea
     {\cal L}_{\rm quadratic}=&&\frac{1}{2}\dot \chi^{\dagger}\dot \chi+\frac{9i}{2\tilde \rho^2}\chi^{\dagger}\dot \chi-\frac{m_H^2}{2}\chi^{\dagger}\chi
     \nonumber\\
     &&+\left(\frac{102}{5\tilde \rho^4}-\frac{37+12\frac{Z^2}{\rho^2}}{192}\right)\chi^{\dagger}\chi\nonumber\\&&+\frac{78i}{5\tilde \rho^2}\chi^{\dagger}\tau^a\chi \chi^a\ .
\eea
Again, this can be interpreted as a system with two harmonic oscillators in a $\rho$ dependent background magnetic field,
coupled with each other by the spin-orbital term. In terms of (\ref{chi12}), we have
\bea
\label{LUNSTABLE}
    {\cal L}=&&\frac{1}{2}(\dot {\vec{x_1}}^2+\dot {\vec{x_2}}^2)+\frac{9}{2\tilde \rho^2}(y_1\dot x_1-x_1\dot y_1+y_2\dot x_2-x_2\dot y_2)\nonumber\\
    &&-\frac{m_H^2+\Omega^2(\rho)}{2}(\vec{x_1}^2+\vec{x_2}^2)+\text{\rm Spin-Orbit} \ .
\eea
with  $\vec{x}_1=(x_1,y_1)$, $\vec{x}_2=(x_2,y_2)$ and
\begin{align}
    \Omega^2(\rho)=-\frac{204}{5\tilde \rho^4}+\frac{37+12\frac{Z^2}{\rho^2}}{96} \ .
\end{align}

We proceed to quantize (\ref{LUNSTABLE}) in the Born-Oppenheimer approximation.
We fix  $y_I$ and $Z$ and first quantize $\vec{x}_1$ and $\vec{x}_2$. This is  justified in the large $m_H$ limit,
where  $\chi$ is fast-moving at frequency $m_H$, while the other degrees of freedom are slow moving with a  typical frequency $\omega_y=\frac{1}{\sqrt{6}}M_{KK}$.

We first look at the $l=0$ state where the spin-orbit coupling vanishes.  In this case $\vec{x}_1$ and $\vec{x}_2$ decouple, and we have two identical harmonic oscillators in the background field
\begin{align}
    \vec{A}=\omega_c(y,-x), \omega_c=\frac{9}{2\tilde \rho^2} \ .
\end{align}
This is the famed Landau problem, with a spectrum
\begin{align}
    E=(n_++\frac{1}{2})\Omega_++(n_-+\frac{1}{2})\Omega_- \ ,
\end{align}
with
\begin{align}
\Omega_{\pm}=\sqrt{m_H^2+\Omega^2+\omega_c^2}\pm \omega_c \ .
\end{align}
At large $m_H$, one has
\begin{align}
    \Omega_{\pm}=m_H \pm \omega_c+\frac{\Omega^2(\rho)+\omega_c^2}{2m_H}+{\cal O}\left(\frac{1}{m_H^2}\right) \ .
\end{align}
Clearly, the $\pm$ solutions can be interpreted as particle/antiparticles. To leading order in ${\cal O}({1}/{m_H})$,
the two frequencies agrees with the case where  $\chi$ is quantized as fermion. Unfortunately,
\begin{align}
\Omega^2(\rho)+\omega_c^2=\frac{81}{4\tilde \rho^4}-\frac{204}{5\tilde \rho^4}<0 \ ,
\end{align}
indicating an instability at the quadratic order. We conclude, that the screening effect in the Coulomb part should not be expanded,
as it causes a charge instability.

\bibliography{HL}

\end{document}